# Structure of Crystalline Water Ice Formed through Neon Matrix Sublimation under Cryogenic and Vacuum Conditions

Reo Sato,[†] So Taniguchi,[†] Naoki Numadate,[†] and Tetsuya Hama[†,*]

[†]*Komaba Institute for Science and Department of Basic Science, The University of Tokyo, 4-6-1 Komaba, Meguro, Tokyo 153-8505, Japan*

AUTHOR INFORMATION

**Corresponding Authors**

*Tetsuya Hama

E-mail: hamatetsuya@g.ecc.u-tokyo.ac.jp

**ABSTRACT**

Ice I has three forms depending on the stacking arrangements of its layers: hexagonal ice $I_h$, cubic ice $I_c$, and stacking disordered ice $I_{sd}$. Below ~60 K, amorphous water becomes metastable, and the formation of any form of ice I is often implicitly precluded. Using a newly developed low-temperature reflection high-energy electron diffraction (RHEED) technique, we show that crystalline ice with cubic stacking sequences (i.e., ice $I_c$) formed through Ne sublimation from a solid $H_2O$/Ne (1:1000 ratio) matrix at 13 K. The extent of staking disorder (disordered cubic and hexagonal stacking sequences) in the ice formed by Ne matrix sublimation is smaller than that in vapor-deposited ice $I_{sd}$ prepared at 143 K and below the limit of detection of low-temperature RHEED. Dependence of the resulting ice structures on the thickness of the $H_2O$/Ne matrix shows that amorphous water first forms in the early stages of Ne sublimation and the cubic stacking sequence subsequently takes place. As the cubic ice $I_c$ formed here at a much lower temperature (13 K) than previously observed (typically above 78 K), Ne matrix sublimation represents a novel route to the formation of cubic ice $I_c$ under low-temperature and low-pressure conditions.

## 1. INTRODUCTION

Water ($H_2O$) ice exhibits rich polymorphism. Twenty one different crystalline structures have been experimentally observed: three forms of ice I (i.e., hexagonal ice $I_h$, cubic ice $I_c$, and stacking disordered ice $I_{sd}$) and ice II–XIX.[1–4] Water is ubiquitous throughout the universe because it comprises hydrogen (H) and oxygen (O), the most and third-most abundant elements, respectively.[5–7] It is an abundant interstellar molecule: ice covers the surface of dust grains in cold (10 K) and dense ($10^4$ molecules $cm^{-3}$) interstellar regions called molecular clouds. These icy dust grains constitute building blocks for the formation of planetary systems, including our solar system.[6–9] The physics and chemistry of water ice at low temperature and low pressure are thus relevant to understanding the structures and properties of icy dust grains in space.[6,7,10–12]

When water molecules cannot diffuse and settle on energetically favorable crystalline sites at low temperatures, metastable amorphous water is generated.[13] For example, vapor deposition of water on a cold substrate below 100 K in a vacuum usually forms amorphous water.[13–20] Infrared (IR) astronomical observations have revealed that ice in molecular clouds is also predominantly amorphous, as evidenced by the lack of a sharp OH-stretching peak around 3230 $cm^{-1}$ (3.1 µm) that is characteristic of crystalline ice.[6–9] In the laboratory, the crystallization of vapor-deposited amorphous water into ice $I_c$ can occur at 129 K on a timescale of ~10,000 s.[11,17,21–24] Jenniskens and Blake reported the activation enthalpy for this transition to be 44 ± 2 kJ $mol^{-1}$.[21,25] The onset of crystallization becomes increasingly later with decreasing temperature,[21,22,25] and below 60 K the timescale would become longer than the age of the universe.[11]

As the structure of water ice strongly depends on the method of its formation, new formation methods can form structures that differ from vapor deposition.[20,26] Hama et al. reported the possibility of crystalline-ice formation at 11–12 K through Ne sublimation from a solid $H_2O$/Ne (1:1000 ratio) matrix.[27] They prepared the Ne matrix at 6 K in a vacuum and subsequently warmed it to 11–12 K. In situ IR reflection–absorption spectroscopy revealed the assembly of the $H_2O$ molecules, which eventually resulted in the formation of ice whose IR spectrum showed the sharp crystalline feature at 3230 cm$^{-1}$.[27] Given that the diffusion of $H_2O$ in the bulk Ne matrix is too slow to attain a crystalline configuration,[28] this low-temperature crystalline ice formation was proposed to occur mainly through the aggregation of water monomers and/or small clusters on the sublimating surface of the solid Ne.[27]

To investigate the crystalline ice structure, Hama et al. also performed experiments using $H_2O$ with 3.5 mol% semiheavy water (HDO) for the measurement of the decoupled OD stretching vibration band; they attributed the appearance of a sharp peak at 2414 cm$^{-1}$ to ice I formed during sublimation of the Ne matrix.[27,29] However, as IR spectroscopy can hardly distinguish among ices $I_h$, $I_c$, and $I_{sd}$,[30–32] diffraction-based analysis is necessary to identify the structure of ice formed by Ne matrix sublimation and facilitate discussion of the mechanism of crystalline ice formation at extremely low temperature. In this study, we developed new experimental apparatus for in situ reflection high-energy electron diffraction (RHEED) at low temperature. Low-temperature RHEED enabled us to elucidate the structure of ice formed by Ne matrix sublimation[33–36] and thus provided insights into its crystallization mechanism. We also discuss the presence of stacking disorder (disordered cubic and hexagonal stacking sequences)[3,4] in the ice formed by Ne matrix sublimation.

## 2. EXPERIMENTAL METHODS

The experimental apparatus comprised an ultrahigh vacuum chamber,[33] an electron gun, and a fluorescent screen (Fig. S1). The chamber was evacuated to ultra-high vacuum conditions (base pressure $10^{-8}$ Pa) using a turbo molecular pump (nEXT300D, Edwards). A mirror-polished aluminum alloy 2017 (Al) substrate (38 mm diameter, HyBridge Co., Ltd.) was mounted on a copper sample holder without removing the native oxide on the Al surface. The copper sample holder was connected to the cold head of a closed-cycle He refrigerator (RDK-101D, Sumitomo Heavy Industries) and installed at the center of the vacuum chamber using a laboratory-made bore-through rotary feedthrough. The temperature of the Al substrate was measured using a silicon diode sensor (DT-670, Lakeshore) placed at the back side of the copper sample holder and controlled at 6 ± 0.2 K using a temperature controller (Model 325, Lakeshore) and a 40 W ceramic heater (MC1010, Sakaguchi E. H Voc Corp.).

Figure 1 outlines the formation of ice by Ne matrix sublimation.[27,37] Purified liquid $H_2O$ (resistivity ≥ 18.2 MΩ cm at 298 K) was obtained from a Millipore Milli-Q water purification system and degassed by several freeze–pump–thaw cycles. A 1:1000 mixture of water vapor and Ne gas (99.999%, Tokyo Gas Chemicals) was introduced onto the Al substrate at 6 K by background deposition, typically for 1800 s at $3.3 \times 10^{-3}$ Pa, measured using a cold cathode gauge (IKR270, Pfeiffer). The actual pressure of the mixed gas was calculated as $1.0 \times 10^{-2}$ Pa, using the gas correction factor for Ne (0.33).[38] The gas flux was estimated as $3.4 \times 10^{16}$ molecules $cm^{-2}$ $s^{-1}$ based on the actual pressure, so deposition for 1800 s corresponded approximately to an exposure of $6.2 \times 10^{19}$ molecules $cm^{-2}$. As there are four molecules in a face-centered cubic unit cell with a lattice parameter of $a = 4.46$ Å,[39] there were ~$2.0 \times 10^{15}$ molecules $cm^{-2}$ on the surface of crystalline Ne. Hence, $6.2 \times 10^{19}$ molecules $cm^{-2}$ correspond to ~31 kilo-monolayers (kML), assuming 1 ML = $2.0 \times 10^{15}$ molecules $cm^{-2}$.

The prepared Ne matrix was warmed at a rate of 0.1 K s$^{-1}$ to 13 ± 1 K to sublimate Ne molecules and form H$_2$O ice with a column density of about 6.2 × 10$^{16}$ molecules cm$^{-2}$. The pressure in the chamber increased by over 10$^{-1}$ Pa during sublimation of the Ne matrix at 13 K. The time required for complete sublimation of Ne at 13 K was 112 ± 10 s, which was counted from when the temperature reached at 13 K (sublimation temperature) to when a sudden pressure decrease down to 10$^{-5}$ Pa was observed. The errors reflect fluctuations in temperature during heating in the independent statistical experiments. Crystalline ice formed here at the Ne sublimation temperature of 13 K, which is in good agreement with Hama et al.'s finding that crystalline ice formed only when Ne sublimated at 11–12 K ±1 K.[27] The condensation rate (flux) of water molecules during the Ne matrix sublimation at 13 K is estimated as 6.2 × 10$^{16}$/112 = 5.5 × 10$^{14}$ molecules cm$^{-2}$ s$^{-1}$. Because this value is lower than a typical flux to form amorphous water by vapor deposition (e.g., 1.8 × 10$^{15}$ molecules cm$^{-2}$ s$^{-1}$ (about 5 × 10$^{-4}$ Pa) at 14 K),[16] the crystalline ice formation at 13 K in this study cannot simply be explained by the latent heat during the aggregation of water molecule.

This study focuses on analysis of the crystalline structure of the ice formed by Ne matrix sublimation at 13 K. The structure of the ice obtained after Ne matrix sublimation was examined in situ by RHEED. An electron beam (20 keV) generated by an electron gun (RDA-004G, R-DEC Co., Ltd.) was impinged on the ice surface. This kinetic energy of electrons corresponds to a de Broglie wavelength $\lambda$ of about 0.09 Å. As the He refrigerator was freely rotatable by the bore-through rotary feedthrough, the incident angle of the electron beam was varied by rotating the Al substrate to maximize the diffraction intensity. Although the incident angle ($\theta$) could not be precisely measured, 2°–3° was desired.[33] Considering the inelastic mean free path of electrons ($L$) in liquid water (about $L$ = 50 nm at 20 keV),[40] the penetration depth of electrons is roughly estimated as 2–3 nm by $L \sin\theta$. Hence, the low-temperature RHEED probes the top several layers of ice samples. The resulting RHEED patterns were projected onto a fluorescent screen and photographed using a digital camera (α7RII, Sony) with a macro lens (SEL50M28, Sony) (Fig. S1). RHEED measurements were performed as quickly as possible (typically within 1 min) to prevent the electron beam damaging the ice samples. We confirmed that no discernible change of RHEED patterns was observed, even after prolonged electron irradiation for 10 min.

We also measured a reference RHEED pattern for ice I prepared by background water vapor deposition: water vapor was first introduced onto an Al substrate at 6 K for 2 min at $8.0 \times 10^{-5}$ Pa to form amorphous water (Fig. S2). The pressure was calculated using the gas correction factor for water (1.25).[38] This corresponds a flux of $2.9 \times 10^{14}$ molecules cm$^{-2}$ s$^{-1}$ and to an exposure of $3.5 \times 10^{16}$ molecules cm$^{-2}$. This exposure corresponds to ~36 ML, assuming 1 ML = $9.9 \times 10^{14}$ molecules cm$^{-2}$ based on a lattice parameter for ice I$_c$ of 6.37 Å.[41,42] Amorphous water was then crystallized into ice I by heating to 143 K and annealing for 10 min (Fig. S2). Prepared vapor-deposited ice I was cooled back to 13 K for RHEED measurements.

## 3. RESULTS

Figure 2(a) shows the RHEED pattern of the H$_2$O/Ne (1:1000) matrix at 6 K just after exposure of ~$6.2 \times 10^{19}$ molecules cm$^{-2}$. The image contrast was dark, but the (111), (200), and (220) diffractions from crystalline Ne were visible (Fig. S3).[43] Figure 2(b) shows the RHEED pattern of the ice obtained after sublimation of the Ne matrix at 13 K. For comparison, the RHEED pattern of vapor-deposited ice I at 13 K is also shown in Fig. 2(c). Both diffraction patterns (Fig. 2(b) and (c)) clearly show Debye rings, indicating the formation of polycrystalline ices. The lower diffraction intensity for the ice obtained after sublimation of the Ne matrix (Fig. 2(b)) than for the vapor-deposited ice I (Fig. 2(c)) indicates the former as containing a minor amount of amorphous water in addition to polycrystalline ice. Hama et al. also reported that both crystalline and amorphous structures coexist in the bulk of the ice obtained after sublimation of the Ne matrix using IR reflection–absorption spectroscopy.[27] To identify the crystalline structures in Fig. 2(b) and (c), we plotted radially integrated diffraction intensity curves with respect to the momentum transfer coordinates, $s$ (Å$^{-1}$):

$$s = \left(\frac{2\pi}{d}\right) = \frac{4\pi}{\lambda}\sin\theta, (1)$$

where $\theta$ and $d$ (Å) are the grazing angle of scattering and the interplanar distance, respectively.[44] Black and gray lines in Fig. 3 show the integrated diffraction intensity curves obtained from the RHEED patterns for ice formed by Ne matrix sublimation (Fig. 2(b)) and vapor-deposition (Fig. 2(c)), respectively. The displayed curves were smoothed by the Savitzky–Golay method. These smoothing treatments do not seriously affect the peak positions, as the widths of the Debye rings are much larger than the picture pixels (Fig. S4).

We first discuss the structure of vapor-deposited ice I (gray line in Fig. 3(a)). The peaks at $d = 2.22$, 1.90, and 1.45 Å agree well with literature values for ice $I_c$ ($d = 2.25$, 1.92, and 1.46 Å, respectively, for the cubic (220), (311), and (331) diffractions) reported in previous X-ray and electron diffraction studies (Table S1).[41,45] Additional small peaks at $d = 2.67$ and 2.04 Å were attributable to hexagonal (102) and (103) diffractions at $d = 2.67$ and 2.07 Å for ice $I_h$ (Table S1).[41,45,46] These results indicate that the vapor-deposited ice I in this study was stacking disordered ice $I_{sd}$; i.e., cubic sequences were partially interlaced with hexagonal sequences. Supporting this, the peak at $d = 3.80$ Å was clearly shifted to a lower value compared with that reported for ice $I_c$ ($d = 3.68$ Å), as shown in Fig. 3(b). Malkin et al. reported that the appearance of a peak around $d = 3.8$ Å is one of the most obvious signatures of stacking disorder.[4] Therefore, we propose that the peak shown by vapor-deposited ice $I_{sd}$ at $d = 3.80$ Å was mainly due to a mixture of hexagonal (100) and cubic (111) diffractions.[33,41,44–46]

The integrated diffraction intensity curve for the ice formed by Ne matrix sublimation has peaks at $d = 2.18$ and $1.87$ Å, which are in good agreement with those for the cubic diffractions observed in the vapor-deposited ice $I_{sd}$ ($d = 2.22$ and $1.90$ Å; Fig. 3(a)). In contrast to the vapor-deposited ice $I_{sd}$ (gray line in Fig. 3(a)), however, the ice formed by Ne matrix sublimation (black line in Fig. 3(a)) did not exhibit clear peaks at around $d = 2.67$ and $2.04$ Å attributable to hexagonal diffractions. These results demonstrate that the ice formed by Ne matrix sublimation at 13 K consisted mainly of cubic stacking sequences. In addition, the peak at $d = 3.67$ Å shown by the ice formed by Ne matrix sublimation was shifted to a lower value compared with that for the vapor-deposited ice $I_{sd}$ ($d = 3.80$ Å) (Fig. 3(b)). The lack of a peak around $d = 3.8$ Å also supports negligible stacking disorder in the ice formed by Ne matrix sublimation.[4]

## 4. DISCUSSION

The present results show that the extent of stacking disorder in the ice formed at 13 K by Ne matrix sublimation was below the detection limit of the low-temperature RHEED experiments, which is apparently smaller than the extent of stacking disorder in ice $I_{sd}$ vapor deposited at 143 K. The formation of ice $I_c$ without stacking disorder has been a topic of intense research. Cubic diffraction patterns have been reported for water ice samples formed by various methods such as vapor deposition on substrates at low temperatures,[3,46,47] crystallization of amorphous water by heating,[47,48] rapid cooling of water droplets,[49] recrystallization of crystalline ices prepared at high pressures,[3,4,50–54] decomposition of clathrate hydrates,[3] and freezing of confined water in porous silica,[55,56] water clusters $(H_2O)_n$ ($n \approx 2000–6000$),[57–59] droplets of supercooled water,[60] and aqueous solutions.[61–63] However, it is becoming increasingly evident that these previous ice samples (except for water clusters) generally show disordered stacking.[3,4,54–56,60–63,46–53] Electron diffraction studies of water clusters have observed only cubic diffractions without hexagonal diffractions.[57–59] Del Rosso et al.[42,64] and Komatsu et al.[65] successfully produced a large amount of pure ice $I_c$ for neutron and X-ray diffraction experiments. Both teams employed a starting material prepared at high pressure: hydrogen-filled ice in the $C_0$ phase prepared above 430 MPa and at 255 K[42,64] and $C_2$ hydrogen hydrate prepared at 3 GPa at room temperature,[65] respectively. In this study, water ice with cubic stacking sequences formed at 13 K, much lower than the crystallization temperatures (78–260 K) typical for the previous studies mentioned above.[3,4,53–62,42,63–65,46–52] The maximum pressure during Ne matrix sublimation (and thus ice formation) was also low ($10^{-1}$ Pa), and there was no need for a starting material prepared at high pressure. Although we do not exclude the possible existence of stacking disorder in the ice formed here by Ne matrix sublimation, the method represents a new route for the low-temperature and low-pressure formation of ice with cubic stacking sequences.

Although clarifying the formation mechanism of ice $I_c$ during Ne matrix sublimation is beyond the scope of this work, we discuss two key factors: the high mobility of $H_2O$ molecules on solid Ne and the size dependence of the stable configuration of water ice. The formation of crystalline ice requires water molecules to find and settle on energetically favorable sites before their motion is stopped by the adsorption of new water molecules. Here, the surface diffusivity of $H_2O$ molecules on solid Ne was much greater than that on water ice, because the van der Waals interactions for the $H_2O$–Ne cluster (0.80 kJ mol$^{-1}$) were much weaker than the hydrogen-bond strength for the $H_2O$–$H_2O$ dimer (20.8 kJ mol$^{-1}$).[66,67] The surface diffusion coefficient of $H_2O$ on solid Ne, $D_s$, can be written as Eq. (2)[13,68,69]:

$$D_s = \nu a_{Ne}^2 \exp(-E/kT), (2)$$

where $\nu$, $a_{Ne}$, $E$, $k$, and $T$ represent the hopping frequency for $H_2O$ ($\nu = 10^{12}$ s$^{-1}$),[66] the lattice parameter of the crystalline Ne ($a_{Ne} = 4.46$ Å),[39] the activation energy of the surface diffusion, Boltzmann's constant, and temperature, respectively. Although the value of $E$ is unknown, adopting $E = 0.80$ kJ mol$^{-1}$ gives $D_s = 1.2 \times 10^{-6}$ cm$^2$ s$^{-1}$ at 13 K. This value is three orders of magnitude larger than the calculated surface diffusion coefficient for $H_2O$ on hexagonal ice at 140 K ($1.0 \times 10^{-9}$ cm$^2$ s$^{-1}$),[66] at which the crystallization of vapor-deposited amorphous water occurs at an observable timescale.[11,17,21–24,70]

The inequality in equation (3) represents the condition for crystalline ice formation during vapor deposition: crystalline ice can form when water molecules diffuse over a larger area than the lattice site area of the crystalline ice within the coverage time, $t_{cover}$, which is the time taken to cover the surface by adsorbed water molecules during vapor deposition.[13]

$$t_{cover} > \frac{a_{ice}^2}{D_s} (3)$$

For Ne matrix sublimation at 13 K, $\frac{a_{ice}^2}{D_s} = 3.4 \times 10^{-9}$ s, as $D_s = 1.2 \times 10^{-6}$ cm$^2$ s$^{-1}$ and the lattice parameter for ice I$_c$ $a_{ice} = 6.37$ Å,[42] and hence $a_{ice}^2 = 4.1 \times 10^{-15}$ cm$^2$. The coverage time $t_{cover}$ can be estimated as $1.8^{+0.1}_{-0.2}$ s, assuming deposition of 63 ML of $H_2O$ molecules ($6.2 \times 10^{16}$ molecules cm$^{-2}$) during the sublimation time of $112 \pm 10$ s at 13 K. This is much greater than $3.4 \times 10^{-9}$ s, and thus the present experiment satisfies the formation condition of crystalline ice.

The above consideration implies that $H_2O$ molecules can attain a stable crystalline configuration upon aggregation following their surface diffusion on the subliming Ne matrix. However, when the initial thickness of the $H_2O$/Ne matrix is decreased to half (an exposure of $3.1 \times 10^{19}$ molecules cm$^{-2}$), amorphous water was found to form at 13 K (Fig. 4(a)). In addition, we confirmed that amorphous water is formed at the sublimation temperature of 14.5 K (Fig. 4(b)). Typical time for Ne sublimation is $53 \pm 1$ s at 14.5 K, which is shorter than that at 13 K ($112 \pm 10$ s). These results are consistent with Hama et al.[27] and indicate that the observed low-temperature crystalline ice formation cannot be explained solely by the high mobility of $H_2O$ on solid Ne.

Both experimental and theoretical studies have shown that the lowest-energy structures are amorphous for small water clusters $(H_2O)_n$ with $n$ below ~100.[71–76] Using a combination of sodium doping and associated IR excitation–modulated photoionization techniques, Moberg et al. recently measured IR spectra for size-selected $(H_2O)_n$ clusters around 150 K generated through homogeneous nucleation in a supersonic expansion of a mixed water–argon gas.[76] They reported that even a water cluster with $n = 90$ can have the ice I structure, whereas water clusters for $n = 90$–150 are mixtures of crystalline and amorphous clusters.[76] The formation condition of ice $I_c$ has also been investigated mainly in homogeneous nucleation systems.[59,77–81] The enthalpy change for the transition from metastable ice $I_c$ to stable ice $I_h$ is small (13–160 J mol$^{-1}$ depending on measurements).[47,82,83] Huang and Bartell deduced the interfacial free energy for the boundary between ice $I_c$ and liquid water to be ~22 mJ m$^{-2}$ at 200 K by experimentally studying water clusters;[59] this is smaller than that between ice $I_h$ and liquid water (about 31 mJ m$^{-2}$).[59,79,84] They thus proposed that the freezing of water to ice $I_c$ is kinetically favorable over ice $I_h$ formation because the lower interfacial free energy associated with ice $I_c$ results in a smaller activation barrier to the liquid, forming a cubic nucleus compared with a hexagonal nucleus.[59] In addition to kinetic effects, Johari theoretically proposed that water droplets of radius smaller than 15 nm and flat films of water thinner than 10 nm freeze to ice $I_c$ at 160–220 K, as ice $I_c$ can be more thermodynamically stable than ice $I_h$ at these nanometer-scales.[79]

In the present study, the aggregation of $H_2O$ molecules mainly occurred on the surface the Ne matrix during sublimation at 13 K; i.e., heterogeneous nucleation.[27] The thickness dependence of the resulting ice structures seen here suggests that the low-energy lattice structure was amorphous, as small water aggregates formed in the early stages of Ne sublimation, and the cubic stacking sequence occurred after the formation of amorphous water (Figs. 2(b) and 4(a)). Although much less is known about heterogeneous systems compared with homogeneous nucleation systems, especially at low temperatures,[85] the thickness dependence observed here is qualitatively in line with the size dependence of the stable configuration of water ice (clusters) in homogeneous nucleation systems.[57–59,71–76,79] This is possibly because the present low-temperature heterogeneous nucleation system involved weak interactions between $H_2O$ and Ne. The formation of amorphous water at 14.5 K suggests that the shorter sublimation time (53 ± 1 s) than 13 K (112 ± 10 s) does not allow the water molecules to assemble into a sufficiently large size of ices (possibly about $n > 100$) in which a crystalline configuration becomes more stable than amorphous water (Fig. 4(b)). More detailed investigation is in progress to elucidate the boundary conditions for the formations of amorphous or crystalline water and the low-temperature crystallization mechanism during Ne matrix sublimation.

AUTHOR INFORMATION

Corresponding Authors

*Tetsuya Hama

Email: hamatetsuya@g.ecc.u-tokyo.ac.jp

ORCIDs

Naoki Numadate: 0000-0001-8992-0812

Tetsuya Hama: 0000-0002-4991-4044

Notes

The authors declare no competing financial interests.

ACKNOWLEDGMENT

This work was supported by JSPS KAKENHI Grant Numbers 21H01143 and 21H05421.

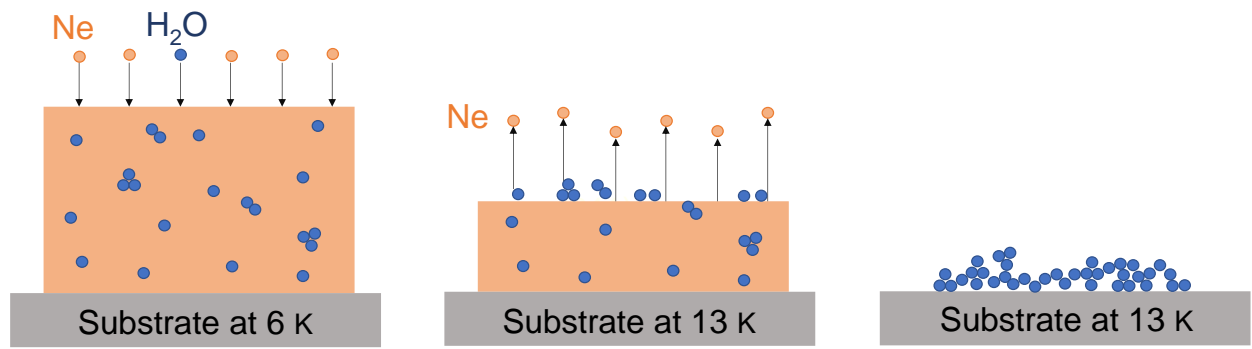

Fig. 1: Schematic illustration of ice formation by Ne matrix sublimation. (left) Deposition of H$_2$O/Ne (1:1000) on an Al substrate at 6 K. (middle) Sublimation of Ne at 13 K. (right) Formation of crystalline ice at 13 K.

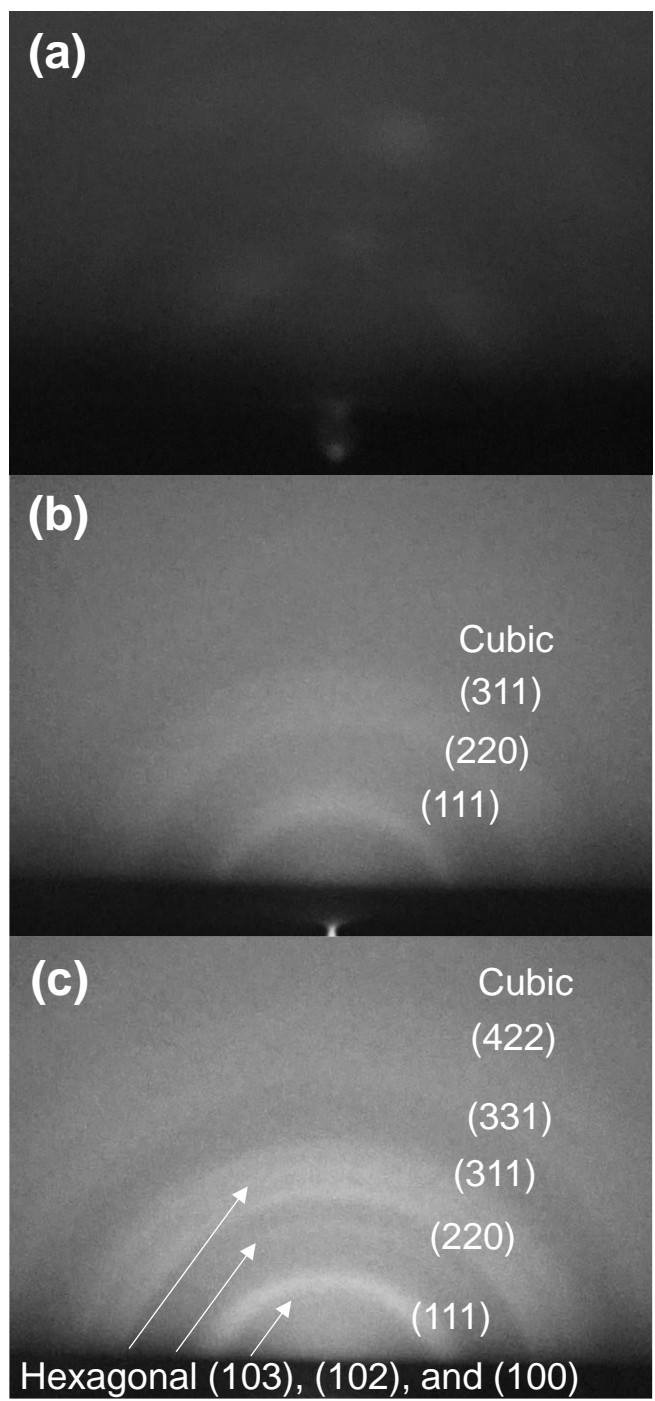

Fig. 2: RHEED patterns of ice obtained by Ne matrix sublimation. (a) A $H_2O$/Ne matrix (1:1000) at 6 K. (b) Ice obtained after Ne matrix sublimation at 13 K. (c) Vapor-deposited ice I at 13 K. Labels (*hkl*) are for the planes of cubic and hexagonal ice (see also Table S1).

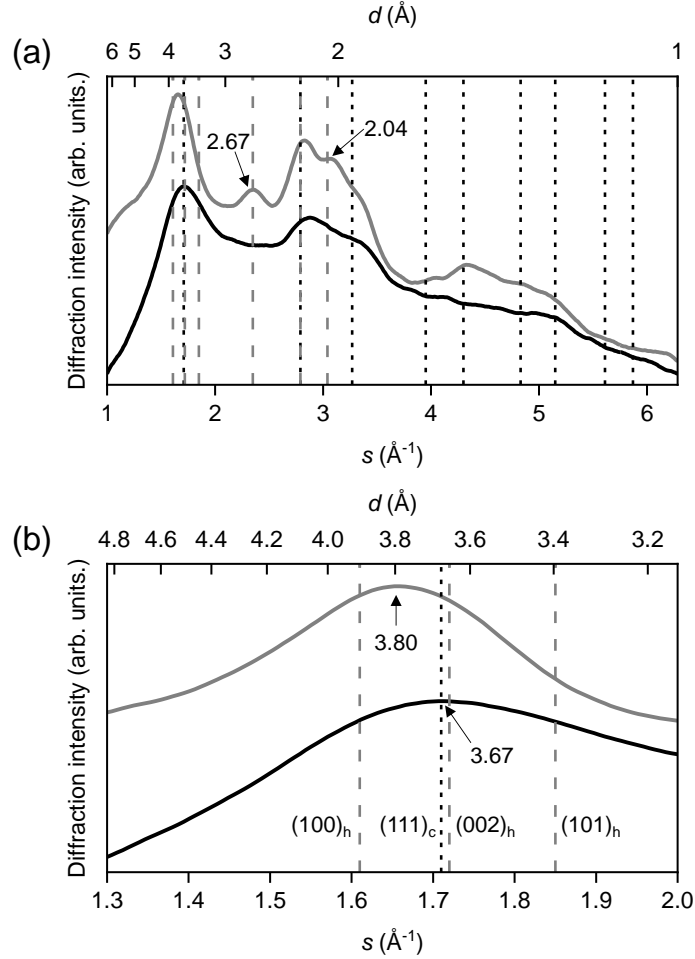

Fig. 3: Integrated diffraction intensity curves for the RHEED patterns of (a) ice obtained after Ne matrix sublimation at 13 K in Fig. 2(b) (black line) and vapor deposited ice I at 13 K in Fig. 2(c) (gray line). Vertical dotted black and dashed gray lines indicate the calculated peak positions for cubic and hexagonal ice, respectively. Table S1 gives details. Black arrows mark the peak positions at $d = 2.67$ and $2.04$ Å, which indicate the (102) and (103) diffractions, respectively, of hexagonal stacking sequences. (b) Magnification of (a) in the range $d = 4.83–3.14$ Å. Black arrows indicate the peak positions at $d = 3.80$ and $3.67$ Å for vapor-deposited ice I and ice obtained after Ne matrix sublimation at 13 K, respectively. Labels (*hkl*) denote the planes of cubic and hexagonal ice (subscript *c* and *h*, respectively).

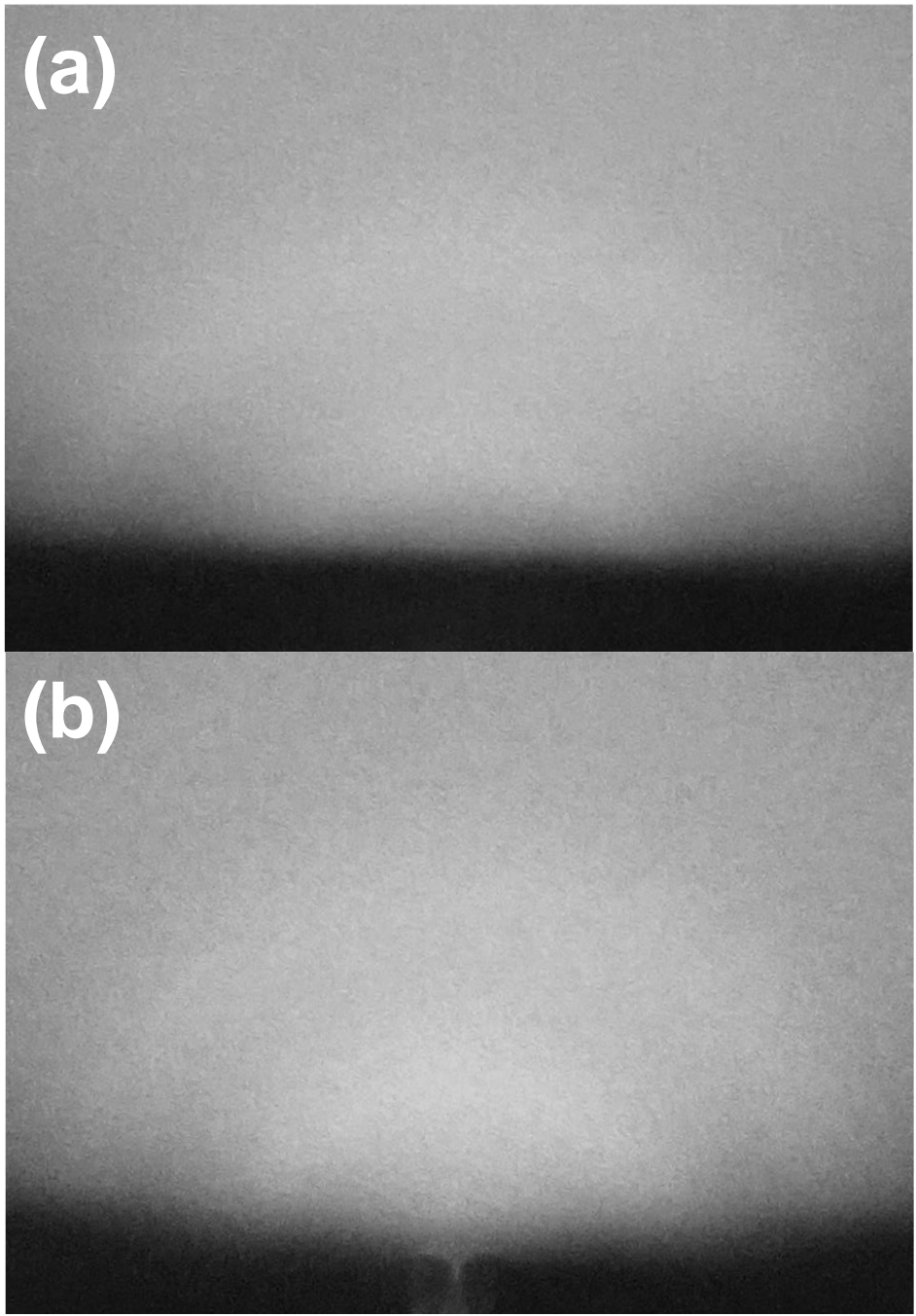

Fig. 4: RHEED patterns of ices obtained by Ne matrix sublimation. (a) Sublimation at 13 K after the H$_2$O/Ne (1/1000) gas exposure of $3.1 \times 10^{19}$ molecules cm$^{-2}$ at 6 K (for 1800 s at $5.0 \times 10^{-3}$ Pa), and (b) Sublimation at 14.5 K after the H$_2$O/Ne (1/1000) gas exposure of $6.2 \times 10^{19}$ molecules cm$^{-2}$ at 6 K (for 1800 s at $1.0 \times 10^{-2}$ Pa). Vague halo patterns indicate the formation of amorphous water in both cases (a) and (b).